\documentclass[review]{elsarticle}
\usepackage{lineno,hyperref,eurosym}
\usepackage{color}
\usepackage{tabularx}
\usepackage[utf8]{inputenc}
\usepackage{graphicx}
\usepackage{amsmath}
\usepackage{lscape}
\usepackage{xcolor, soul}
\usepackage{makecell}

\journal{Acta Astronautica}
\begin{document}
\begin{frontmatter}
\title{Design of a low-thrust gravity-assisted rendezvous trajectory to Halley's comet}
\author[label1]{R. Flores}
\author[label1]{A. Beolchi}
\author[label1]{E. Fantino$^{\star}$}
\author[label1]{C. Pozzi}
\author[label2]{M. Pontani}
\author[label3]{I. Bertini}
\author[label4,label5]{C. Barbieri}
\address[label1]{Department of Aerospace Engineering, Khalifa University of Science and Technology, 
P.O. Box 127788, Abu Dhabi (United Arab Emirates)}
\address[label2]{Department of Astronautical, Electrical, and Energy Engineering, Sapienza Universit\`a di Roma, Rome (Italy)}
\address[label3]{Department of Science and Technology, University of Naples "Parthenope", Naples (Italy)}
\address[label4]{Department of Physics and Astronomy "G.
Galilei", University of Padova, Padua (Italy)}
\address[label5]{INAF, Istituto Nazionale di Astrofisica (Italy)}
\cortext[cor]{Corresponding author: elena.fantino@ku.ac.ae (E.~Fantino)}

\begin{abstract}
Comets are the most pristine planetesimals left from the formation of the Solar System. They carry unique information on the materials and the physical processes which led to the presence of planets and moons. Many important questions about cometary physics, such as origin, constituents and mechanism of cometary activity, remain unanswered. The next perihelion of comet 1P/Halley, in 2061, is an excellent opportunity to revisit this object of outstanding scientific and cultural relevance. In 1986, during its latest approach to the Sun, several flyby missions from Europe, Soviet Union, and Japan targeted Halley's comet to observe its nucleus and shed light on its properties, origin, and evolution. However, due to its retrograde orbit and high ecliptic inclination, the quality of data was limited by the large relative velocity and short time spent by the spacecraft inside the coma of the comet. A rendezvous mission like ESA/Rosetta would overcome such limitations, but the trajectory design is extremely challenging due to the shortcomings of current propulsion technology. Given the considerable lead times of spacecraft development and the long duration of the interplanetary transfer required to reach the comet, it is imperative to start mission planning several decades in advance. This study presents a low-thrust rendezvous strategy to reach the comet before the phase of intense activity during the close approach to the Sun. The trajectory design combines a gravity-assist maneuver with electric propulsion arcs to maximize scientific payload mass while constraining transfer duration. A propulsive plane change maneuver would be prohibitive. To keep the propellant budget within reasonable limits, most of the plane change maneuver is achieved via either a Jupiter or a Saturn flyby. The interplanetary low-thrust gravity-assisted trajectory design strategy is described, followed by the presentation of multiple proof-of-concept solutions.
\end{abstract}

\begin{keyword}
1P/Halley \sep Trajectory design \sep Trajectory optimization  \sep Cometary  rendezvous  \sep Low thrust  \sep Gravity assist
\end{keyword}

\end{frontmatter}


\section{Introduction}
\label{sec_introduction}
Comets are not only objects that have long captivated the interest of the general public but also celestial bodies of great scientific importance. Understanding where and how comets formed and reached their present reservoirs, the nature of their constituent materials, and the mechanics of their activity is essential to determining the role these bodies played in planetary formation and even in the appearance of water and life on Earth \cite{Barbieri:2017}.
Comets are categorized into short and long period, with the boundary between the two groups set at an orbital period of 200 years. Compared to the long-period objects, many short-period comets are fainter, less active, and often lack distinctive features, such as the characteristic tail. Nonetheless, short-period comets are the only ones whose appearance can be predicted accurately, making the members of this subset the ideal targets of cometary missions. 

1P/Halley stands out among short-period comets due to its cultural and scientific relevance. It is large and bright, which facilitated naked-eye sightings, observations and records during at least two millennia. Therefore, its orbit is wkown with high precision, which facilitated planning for the space missions in 1985/1986  \cite{Brady:1971,Yeomans:1981}. Its next perihelion, in 2061, is expected to attract significant attention from space agencies worldwide, prompting the design and realization of space missions to visit it. However, Halley's comet presents several challenges for mission planning due to its retrograde orbit and high ecliptic inclination (162 degrees) and eccentricity (0.967), which disfavor prograde intercepts. 

During its latest approach to the Sun in 1986, several space missions encountered 1P/Halley at relative velocities of 70-80 km/s \cite{Reinhard:1982,Stelzried:1986,Oya:1986,Sagdeev:1986,Keller:1986,Reinhard:1987,Akiba:1980}. This resulted in a brief period of close observation within the coma, which limited both the quality and quantity of the collected data \cite{Reinhard:1986,Curdt:1988,Volwerk:2014,Keller:2020}.  
Only one hemisphere of the nucleus could be imaged, and the rotational state could not be accurately measured. Scientists are tempted to explain the nucleus of 1P/Halley as a contact binary, but only an extended collection of high-resolution images covering the entire body can confirm this hypothesis. Several other matters require detailed investigation, such as the geomorphology of the surface, the localization and dimension of the active regions, the amount of mass loss, and the internal structure. To address most of these unresolved questions, it is crucial to encounter Halley's comet well before the onset of its activity, i.e., at a heliocentric distance larger than the radius of Mars' orbit.

The above considerations, explicated in \cite{Barbieri:2025}, justify the realization of a rendezvous mission, one in which the spacecraft approaches the comet at zero relative velocity. The design of a trajectory  of this kind is challenging, though, primarily due to the prohibitive cost to reverse the direction of the spacecraft's heliocentric motion and match the energy of the orbit of the comet. Another complication is that the major axis of 1P/Halley is almost perpendicular to its line of nodes. For many short-period comets, these two directions are nearly aligned, making it possible to perform rendezvous via a three-impulse ballistic trajectory, characterized by a single mid-course maneuver near the line of nodes \cite{Friedlander:1971}. A rendezvous trajectory to Halley's comet cannot benefit from the three-impulse scheme.

In 1971, Friedlander et al. \cite{Friedlander:1971} placed 1P/Halley in third position of a list of interesting targets for comet rendezvous missions in the period 1975-1995, only preceded by the much more accessible comets Encke and Kopff. However, the authors in \cite{Friedlander:1971} believed that the practical realization of a rendezvous mission to Halley's comet was tied to the development of nuclear propulsion, a remark that is shared by many other contributions of the time and more recently by Horsewood et al. \cite{Horsewood:2023}. Nuclear engines are the preferred propulsion system in most 1P/Halley rendezvous studies prior to 1986 \cite{Kruse:1969,Friedlander:1970,Friedlander:1971b} because this technology was expected to be soon available, thus solving all the dynamical complications related to the mission and simultaneously enabling large payload mass, low characteristic launch energy, short flight time (a key constraint due to the late start of the research) and  wide launch windows. An alternative to direct trajectories that employ nuclear propulsion consists in incorporating a substantial gravity assist (GA) maneuver. The idea of using the gravitational field of a giant planet to facilitate the rendezvous with Halley's comet dates back to 1967 \cite{Michielsen:1968}. Also at that time, reaching the comet did not seem so far fetched, and  between the late 60s and the late 70s notable efforts were devoted to designing possible trajectories \cite{Friedlander:1967,Deerwester:1969,Manning:1970,Farquhar:1977,Farquhar:1978,Jacobson:1978,West:1978,Farquhar:1979,Farquhar:1982}. However, with the exception of Michielsen \cite{Michielsen:1968}, who believed that a rendezvous with comet Halley with realistic propulsive capabilities required the assistance of the planets, gravity-assisted trajectories were always regarded secondary to mission concepts based on nuclear propulsion. Even more importantly, no previous research investigates the joint use of unpowered GA and continuous thrust arcs, with the exception of the works of Horsewood et al. \cite{Horsewood:2023} and Ozaki et al. \cite{Ozaki:2020} which are based on high-power (40-45 kW) and high-thrust (368-2250 mN) nuclear and solar-electric propulsion, respectively. Once more, the reason for the relatively little importance awarded to gravity-assisted trajectories is to be found in the late start of the studies, and the relatively long times (approximately 10 years \cite{Friedlander:1971}) expected for the development of nuclear engines. These two factors made it so that most analyses could only consider launch dates after 1980, when Jupiter and Saturn were not in favorable positions for a GA maneuver, and the resulting short permissible time of flight implied that only nuclear propulsion could single-handedly perform the required maneuvers.

These previous investigations explain the importance of starting mission planning decades in advance in order to properly assess the feasibility of the mission with readily-available propulsion technology. The next perihelion of 1P/Halley is expected in 2061, and therefore it is crucial to investigate rendezvous opportunities at once. The present contribution proposes a design strategy combining low-thrust (LT) arcs and planetary GA. The thrust is provided by a 36 mN Hall effect motor drawing 640 W of electrical power from radioisotope thermoelectric generators (RTGs). In this way, the performance characteristics assumed in the analysis are attainable with existing hardware.

This manuscript is organized as follows. Section~\ref{sec_dyn_mod} defines the mathematical model and provides the relevant planetary and spacecraft data. The trajectory design strategy is the subject of Sect.~\ref{sec_tra_des}. The numerical results are presented and discussed in Sect.~\ref{sec_res_dis}. Finally, Sect.~\ref{sec_con} is devoted to the concluding remarks. This article improves on a previous work presented at the 75$^{\rm th}$ International Astronautical Congress \cite{Beolchi:2024}.

\section{Dynamical model}
\label{sec_dyn_mod}
The ballistic trajectories are computed with a simple patched conics model. The motion of the spacecraft during each segment is governed by the gravity of a single celestial body: the Sun for heliocentric arc and the relevant planet (Jupiter or Saturn) during the GA maneuver. The flyby is considered instantaneous, assuming a zero-size Sphere Of Influence (SOI). Furthermore, the orbits of the planets are approximated by coplanar circles with the Sun at their center. This simplified model is sufficient for the purpose of demonstrating the feasibility of the mission concept.

The orbital elements of the planets and Halley's comet, as well as the values of relevant gravitational parameters, have been downloaded from JPL's Solar System Dynamics website \cite{horsys,spicenaif}. All the elements are osculated at the epoch of 1P/Halley perihelion (28 July 2061) and referenced to the mean ecliptic and equinox of J2000 reference frame. For the planets, the eccentricity and inclination are set to zero, following the coplanar circular orbits assumption. Table~\ref{tab_hal} list the orbital elements of the comet (semimajor axis $a$, eccentricity $e$, inclination $i$, longitude of the ascending node $\Omega$, argument of pericenter $\omega$ and epoch of pericenter $t_p$). 

\begin{table}[h!]
\centering
\caption{Orbital elements of 1P/Halley in ecliptic frame.}
\begin{tabular}{|r|r|r|r|r|r|}
\hline
$a$ & $e$ & $i$ & $\Omega$ & $\omega$  & $t_p$ \\
\hline
17.7 au & 0.96 & $162^\circ$ & $59.4^\circ$ & $112^\circ$ & JD 2474034.2 \\ \hline
\end{tabular}
\label{tab_hal}
\end{table}

The rendezvous mass of the spacecraft is set to 1000 kg, intermediate between the dry mass of ROSETTA (1180 kg, including 165 kg of scientific payload and 100 kg of lander \cite{Glassmeier:2007}) and that of Deep Impact (EPOXI) (879 kg, including 364 kg of impactor payload \cite{Blume:2005}). The propulsion system is assumed to be fully steerable, delivering constant specific impulse and maximum thrust. Because the spacecraft travels toward the outer Solar System and remains far from the Sun for most of the mission duration, RTGs are chosen as the source of electric power. The spacecraft and thruster data (thrust $T$, specific impulse $I_{sp}$ and input power $P_{in}$) are listed in Table~\ref{tab:eng}. The engine parameters correspond to one of three settings of the PPS\textcopyright X00 Hall Effect Thruster (HET) \cite{Vaudolon:2019} used by Fantino et al. to design a tour of the inner moons of Saturn \cite{Fantino:2023}. While other types of electric motors deliver higher specific impulse, HETs have lower power requirement for a given thrust. This is a critical advantage, in view of the power constraints.

\begin{table}[h!]
\centering
\caption{Assumed thruster performance parameters.}
\begin{tabular}{|r|r|r|r|} \hline
$T$ & $I_{sp}$ & $P_{in}$ \\ \hline
36 mN & 1600 s & 640 W \\ \hline
\end{tabular}
\label{tab:eng}
\end{table}

\section{Trajectory design method}
\label{sec_tra_des}
This work adopts a two-step strategy. Initially, a set of traditional impulsive trajectories is computed, minimizing the total impulse required for the transfer between the flyby planet and the comet. Next, the impulsive solution is transformed into continuous LT arcs, compatible with the use of electric propulsion.

The trajectory of the spacecraft is divided into three stages: 
\begin{itemize}
\item Transfer between Earth and flyby planet.
\item GA with either Jupiter or Saturn.
\item Transfer between flyby planet and 1P/Halley. 
\end{itemize}

In the following, subscript $1$ denotes the departure event, $2$ is the state immediately before the GA (i.e., arrival at the flyby planet), $3$ is the condition right after the flyby, and $4$ is the rendezvous with the comet. For example, due to the instantaneous GA with zero-radius SOI hypothesis $t_2=t_3$ and ${\bf r}_2={\bf r}_3$ (time and position are continuous) but ${\bf v}_2\neq{\bf v}_3$ (velocity is discontinuous).

Additionally, superscript $S$ denotes the Sun, $E$ stands for Earth, $F$ is the flyby planet and $H$ represents 1P/Halley. Because the spacecraft must rendezvous with the comet, ${\bf r}_4={\bf r}^H_4$ and ${\bf v}_4={\bf v}^H_4$.

A preliminary study \cite{Beolchi:2024} determined that the departure energy from Earth required to lower the planet-to-comet transfer impulse to reasonable values (i.e., compatible with the performance of the propulsion system) is very high. Thus, to maximize utilization of the launch vehicle's energy, the departure from Earth is chosen to coincide with the perihelion of the pre-GA conic (i.e., ${\bf v}_1$ is parallel to ${\bf v}^E_1$).

\subsection{Step 1: Impulsive trajectory design}
\label{sec:imp_tra}
All calculations, except the GA maneuver, use a heliocentric Cartesian ecliptic reference frame $xyz$ with the unit vector $\bf i$ pointing along the direction of the spring equinox, $\bf k$ along the normal to the ecliptic plane, and $\bf j$ completing the right-handed triad \{${\bf i},{\bf j},{\bf k}$\}.

The time of flyby ($t_2$) is selected as the independent variable. It determines completely the feasible departure conditions from Earth, reducing the optimization problem to finding the most advantageous time of rendezvous with the comet ($t_4$). This is possible because the optimal flyby parameters (depth and inclination of the planetocentric hyperbola) can be computed directly from $t_2$ and $t_4$, thus reducing the dimensionality of the problem.

\subsubsection{Earth-GA segment}
\label{seg_12}
For a given flyby date, the ecliptic longitude of the GA planet is computed from the ephemeris as
\begin{equation}
	l^F_2 = \Omega^F + \omega^F + n^F (t_2 - t^F_p),
\end{equation}
where $n^F$ is the mean motion of the planet.
The minimum eccentricity of the conics that connect Earth with the planet is determined by the condition of being able to reach the planet:
\begin{equation}
	a_{\min} = \frac{r^E + r^F}{2} \to e_{\min} = 1 - \frac{r^E}{a_{\min}}.
\end{equation}
The upper bound of eccentricity is set by the maximum characteristic energy ($C3$) that the launcher is capable of:
\begin{equation}
	v_{1,{\max}} = \sqrt{C3} + v^E = \sqrt{C3} + \sqrt {\frac{\mu^S}{r^E}},
\end{equation}
\begin{equation}
	E_{\max} = \frac{v^2_{1,{\max}}}{2} - \frac{\mu^S}{r^E} \to a_{\max} = \frac{-\mu^S}{2E_{\max}} \to e_{max} = 1 - \frac{r^E}{a_{\max}},
\end{equation}
where $E$ denotes the mechanical energy and $\mu$ is the gravitational parameter.
Due to the condition of departure from the perihelion, the eccentricity determines the transfer angle between Earth and the planet
\begin{equation}
	\cos(\theta_{12}) = \frac{1}{e} \left( \frac{p}{r^F} - 1 \right),
\end{equation}
with $p=a(1-e^2)$ denoting the semilatus rectum. Likewise, the transfer time $t_{12}$ (only the elliptic case is shown for the sake of brevity) can be computed from $e$
\begin{equation}
	cos(E) = \frac{e+\cos(\theta_{12})}{1+e\cos(\theta_{12})} \to M=E-e\sin(E) \to t_{12}=M \sqrt{\frac{a^3}{\mu^S}},
\end{equation}
where $E$ and $M$ are the eccentric and mean anomalies, respectively.
The departure time is therefore
\begin{equation}
	t_1 = t_2 - t_{12}(e),
\end{equation}
and the departure ecliptic longitude
\begin{equation}
	l_1 = l^F_2 - \theta_{12}(e).
\end{equation}
The feasible trajectories are those where $l_1$ coincides with the position of Earth at $t_1$:
\begin{equation} \label{eq_e}
	l^F_2 - \theta_{12}(e) = \Omega^E + \omega^E + n^E (t_2 - t_{12}(e) - t^E_p).
\end{equation}
This nonlinear equation is solved numerically with the secant method, yielding the values of eccentricity that allow the spacecraft to reach the planet at $t_2$. Usually, more than one solution exists.

Once the eccentricity is known, the spacecraft state just before the GA is easy to determine.
\begin{equation}
	a = \frac{r^E}{1-e} \to E = \frac{-\mu^S}{a} \to v^2_1 = 2 \left( E + \frac{\mu^S}{r^E} \right) \; ; \; v^2_2 = 2 \left( E + \frac{\mu^S}{r^F} \right) \; ,
\end{equation}
\begin{equation}
	v_{2,\theta} = v_1 \frac{r^E}{r^F} \; ; \; v_{2,r} = \sqrt{v^2_2 - v^2_{2,\theta}} \; ,
\end{equation}
\begin{equation} \label{eq_v2}
	{\bf r}_2  = r^F {\bf u}_r \; ; \;  {\bf v}_2  = v_{2,r} {\bf u}_r +  v_{2,\theta} {\bf u}_\theta \; .
\end{equation}
where ${\bf u}_r = \cos(l^F_2) {\bf i} + \sin(l^F_2) {\bf j}$ and ${\bf u}_\theta = {\bf k} \times {\bf u}_r$.

\subsubsection{GA-Halley segment}
\label{seg_34}
Before tackling the GA, the last segment of the trajectory is discussed. It serves to define a parameter required for computing the optimal flyby configuration.

For each pair of values $\{t_2,t_4\}$ a Lambert problem is defined that connects the points ${\bf r}_3={\bf r}_2$ and ${\bf r}_4 = {\bf r}^H(t_4)$ with a time of flight $TOF=t_4-t_3$. Let ${\bf v}^L_3$ and ${\bf v}^L_4$ denote the terminal velocities of the Lambert arc. The two impulses required to rendezvous with the comet using a ballistic trajectory are $\Delta V_3 = \| {\bf v}^L_3 - {\bf v}_3 \|$ and $\Delta V_4 = \| {\bf v}^H_4 - {\bf v}^L_4 \|$. The next subsection presents an analytical method to determine the optimal value of ${\bf v}_3$. For the time being, it will be assumed that ${\bf v}_3$ is known.

The optimal rendezvous trajectory (for a given $t_2$) is determined minimizing the total impulse $\Delta V_{\textrm{tot}} = \Delta V_3 + \Delta V_4$ with respect to $t_4$. This is achieved using Brent's method, which combines the golden section search with parabolic interpolation \cite{Brent:2002}.
The search space is constrained to rendezvous dates between the comet's crossing of the orbits of Uranus (June 2053) and Mars (May 2061). Trajectories that intercept the comet farther from the Sun than Uranus are too short, in the TOF sense, for the propulsion system to provide the necessary impulse. Reaching the comet outside the orbit of Mars is a scientific requirement. It ensures that the beginning of the high-activity phase can be observed.

\subsubsection{Determination of the optimal flyby parameters}
\label{sss:flyby}
For a given pair $\{t_2,t_4\}$, the arrival velocity to the flyby planet ${\bf v}_2$ and the first Lambert terminal velocity ${\bf v}^L_3$ are known. The objective is to find the combination of deflection angle $\delta \in [0,\pi[$ of the relative velocity, and inclination of the planetocentric hyperbola $\alpha \in [0,2\pi[$, that minimizes $\Delta V_3$. The deflection is related to the pericenter radius ($r_\pi$) through \cite{Broucke:1988}
\begin{equation} \label{eq_deltamax}
	\sin\left( \frac{\delta}{2} \right) = \frac{1} {1 + \frac{r_{\pi} v^2_\infty}{\mu^F}},
\end{equation}
where $v_\infty$ denotes the magnitude of the inbound relative velocity ${\bf b}_2 = {\bf v}_2 - {\bf v}^F_2$. The post-GA velocity is obtained from ${\bf v}_3 = {\bf b}_3 + {\bf v}^F_3$ (note that ${\bf v}^F_3 = {\bf v}^F_2$), where the angle between ${\bf b}_2$ and ${\bf b}_3$ is $\delta$ and both have the same magnitude. Hereafter, to reduce clutter in the notation, the planet's velocity during the flyby (the only time when it is relevant) will be written simply as ${\bf v}^F$. 

\begin{figure}[h!]
\centering
\includegraphics[width=0.40\textwidth]{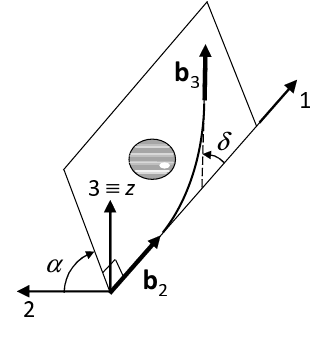}
\caption{Flyby geometry in planetocentric frame $\{ {\bf u}_1 , {\bf u}_2 , {\bf u}_3 \}$.}
\label{fig:PlanetRefFrame}
\end{figure}

Let us define a planetocentric reference frame with directions $\{ {\bf u}_1 , {\bf u}_2 , {\bf u}_3 \}$, where $ {\bf u}_1 = \frac{{\bf b}_2}{v_\infty}$ (i.e., the direction of the inbound relative velocity), ${\bf u}_3 = {\bf k}$ and  ${\bf u}_2 = {\bf k} \times {\bf u}_1$ (see Fig.~\ref{fig:PlanetRefFrame}).
In this system, the components of ${\bf v}_3$ are
\begin{equation}
	(v_\infty\cos\delta + v^F_1 , v_\infty\sin\delta\cos\alpha  + v^F_2 , v_\infty\sin\delta\sin\alpha),
\end{equation}
where $v^F_i = {\bf v}^F \cdot {\bf u}_i$. Note that $\alpha = 0$ corresponds to a trailing flyby in the plane of the ecliptic, $\alpha = \pi/2$ is a GA perpendicular to the ecliptic approaching the planet from the South, $\alpha = \pi$ is an in-plane leading flyby, and $\alpha = 3\pi/2$ is perpendicular to the ecliptic approaching from the North.
Let ${\bf d} = {\bf v}^F - {\bf v}^L_3$; the components of ${\bf v}_3 - {\bf v}^L_3$ in the $123$ system become:
\begin{equation} \label{eq_v3mvl}
	(v_\infty\cos\delta + d_1 , v_\infty\sin\delta\cos\alpha  + d_2 , v_\infty\sin\delta\sin\alpha + d_3).
\end{equation}
The post-GA impulse magnitude can be computed from
\begin{equation} \label{eq_dv32a}
	\Delta V^2_3 = ({\bf v}_3 - {\bf v}^L_3) \cdot ({\bf v}_3 - {\bf v}^L_3).
\end{equation}
Inserting Eq.~\ref{eq_v3mvl} into Eq.~\ref{eq_dv32a} yields, after some tedious but straightforward algebra, 
\begin{equation} \label{eq_dv32b}
	\Delta V^2_3 = K + 2v_\infty (d_1\cos\delta + d_2\sin\delta\cos\alpha  + d_3\sin\delta\sin\alpha),
\end{equation}
with $K = v^2_\infty + d^2$. The flyby parameters $\{\delta$, $\alpha\}$ that minimize $\Delta V_3$ can be determined computing the derivatives of Eq.~\ref{eq_dv32b}.
\begin{equation} \label{eq_der1del}
	\frac{1}{2v_\infty} \frac{\partial \Delta V^2_3}{\partial \delta} = \cos\delta (d_2\cos\alpha + d_3\sin\alpha) - d_1\sin\delta,
\end{equation}
\begin{equation} \label{eq_der1i}
		\frac{1}{2v_\infty} \frac{\partial \Delta V^2_3}{\partial \alpha} = \sin\delta (d_3\cos\alpha - d_2\sin\alpha),
\end{equation}
\begin{equation} \label{eq_der2del}
	\frac{1}{2v_\infty} \frac{\partial^2 \Delta V^2_3}{\partial \delta^2} = -\sin\delta (d_2\cos\alpha + d_3\sin\alpha) - d_1\cos\delta,
\end{equation}
\begin{equation} \label{eq_der2i}
	\frac{1}{2v_\infty} \frac{\partial^2 \Delta V^2_3}{\partial \alpha^2} = -\sin\delta (d_3\sin\alpha + d_2\cos\alpha).
\end{equation}
The optimal inclination is obtained setting Eq.~\ref{eq_der1i} to zero. This yields three possible solutions, one is trivial ($\delta=0$, meaning no GA), and the other two (spaced $\pi$ apart) are given by
\begin{equation} \label{eq_iopt}
	\tan\alpha = \frac{d_3}{d_2}.
\end{equation}
One solution corresponds to a local maximum of $\Delta V_3$ and must be discarded. The optimal inclination $\alpha_{opt}$ makes Eq.~\ref{eq_der2i} positive. Because $\delta \in [0,\pi[$, $\sin\delta$ is positive, therefore:
\begin{equation} \label{eq_cond_alpmin}
	d_3\sin\alpha_{opt} + d_2\cos\alpha_{opt} < 0.
\end{equation}
Next, the optimal deflection $\delta_{opt}$ is computed from Eq.~\ref{eq_der1del}
\begin{equation} \label{eq_delopt}
	\tan\delta = \frac{d_2\cos\alpha_{opt} + d_3\sin\alpha_{opt}}{d_1},
\end{equation}
Only one solution of Eq.~\ref{eq_delopt} is relevant, because $\delta \in [0,\pi[$. To verify that this solution is indeed a minimum, use Eq.~\ref{eq_der2del}, keeping in mind that $\sin\delta$ is positive:
\begin{equation} \label{eq_cond_delmin}
	d_2\cos\alpha_{opt} + d_3\sin\alpha_{opt} + d_1\cot\delta_{opt} < 0.
\end{equation}
Substituting Eq.~\ref{eq_delopt} into Eq.~\ref{eq_cond_delmin} gives
\begin{equation} \label{eq_cond_delmin_2}
	\frac {d^2_1 + (d_2\cos\alpha_{opt} + d_3\sin\alpha_{opt})^2} {d_2\cos\alpha_{opt} + d_3\sin\alpha_{opt}} < 0.
\end{equation}
Because the numerator is always positive, and the denominator is negative by virtue of Eq.~\ref{eq_cond_alpmin}, the inequality is automatically true and the extremum is always a minimum of $\Delta V_3$.

The optimum solution is only feasible if $\delta_{opt} \leq \delta_{max}$, where $\delta_{max}$ is obtained inserting the minimum acceptable flyby pericenter radius in Eq.~\ref{eq_deltamax}. If $\delta_{opt}$ is too large, the absolute minimum must lie on the boundary of the $[0,\delta_{max}]$ interval. In this case, use $\delta = \delta_{max}$ because the minimum cannot be at $\delta = 0$, that would require a local maximum inside $]0,\delta_{opt}[$.

The procedure outlined above determines the optimal GA geometry directly from the pair $\{t_2,t_4\}$. Therefore, for a given flyby date, the only free parameter is the rendezvous date. Because the dimensionality of the optimization problem has been reduced to 1, the computation is extremely efficient. 

\subsection{Low-thrust trajectory design}
\label{sec_ltt} 
The impulsive maneuver at rendezvous $\Delta V_4$ and the post-flyby impulse $\Delta V_3$ can be eliminated by introducing one or more LT arcs in the post-flyby segment.
The method applied in this work preserves the event dates (flyby date $t_2$ and rendezvous date $t_4$) and the post-flyby spacecraft state determined in the impulsive solution (${\bf r}_3,{\bf v}_3$). The LT transfer is built propagating the trajectory backwards in time from the rendezvous state and forward from the post-GA conditions. The direction of thrust at each step is chosen by minimizing the impulse required to connect the backward and forward branches with a ballistic arc.
The effect of the propulsion system is approximated by a series of small instantaneous impulses, connected with short Keplerian arcs. Let $\Delta {\bf v}_i$ denote the discrete impulses ($i = 1,...,n$). They are linked by $n-1$ ballistic arcs of equal heliocentric amplitude $\Delta \theta$. The magnitude of each impulse is given by the product of the instantaneous acceleration (computed from the thrust, assumed constant, and spacecraft mass) and the transfer time of the corresponding arc $\Delta t_i$
\begin{equation}
\Delta t_i = \displaystyle \frac{r_i^2 \Delta \theta}{h_i},
\end{equation} 
with $h_i$ and $r_i$ the instantaneous values of the angular momentum and heliocentric distance.

\begin{figure}[h!]
\centering
\includegraphics[width=0.56\textwidth]{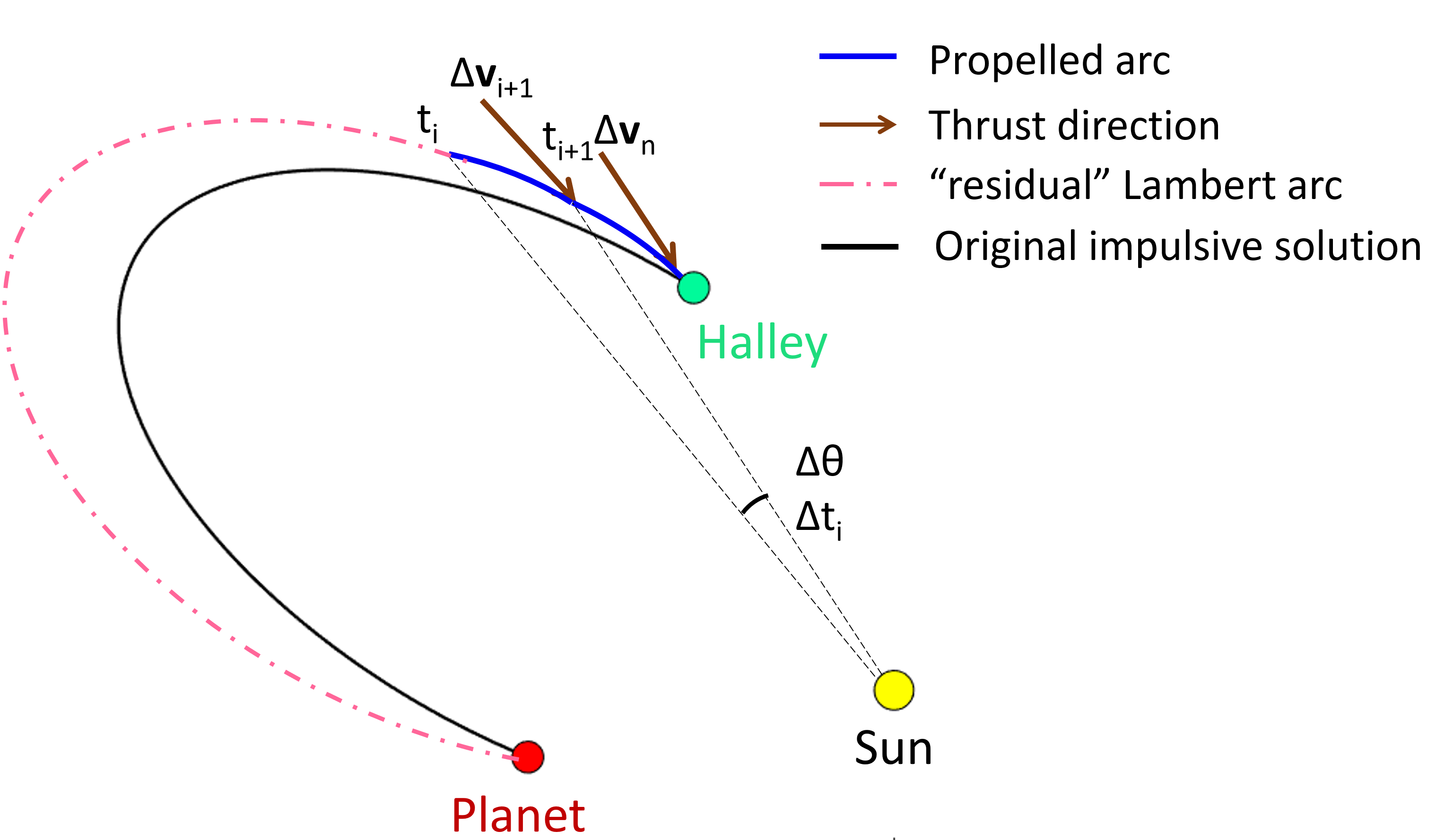}
\includegraphics[width=0.37\textwidth]{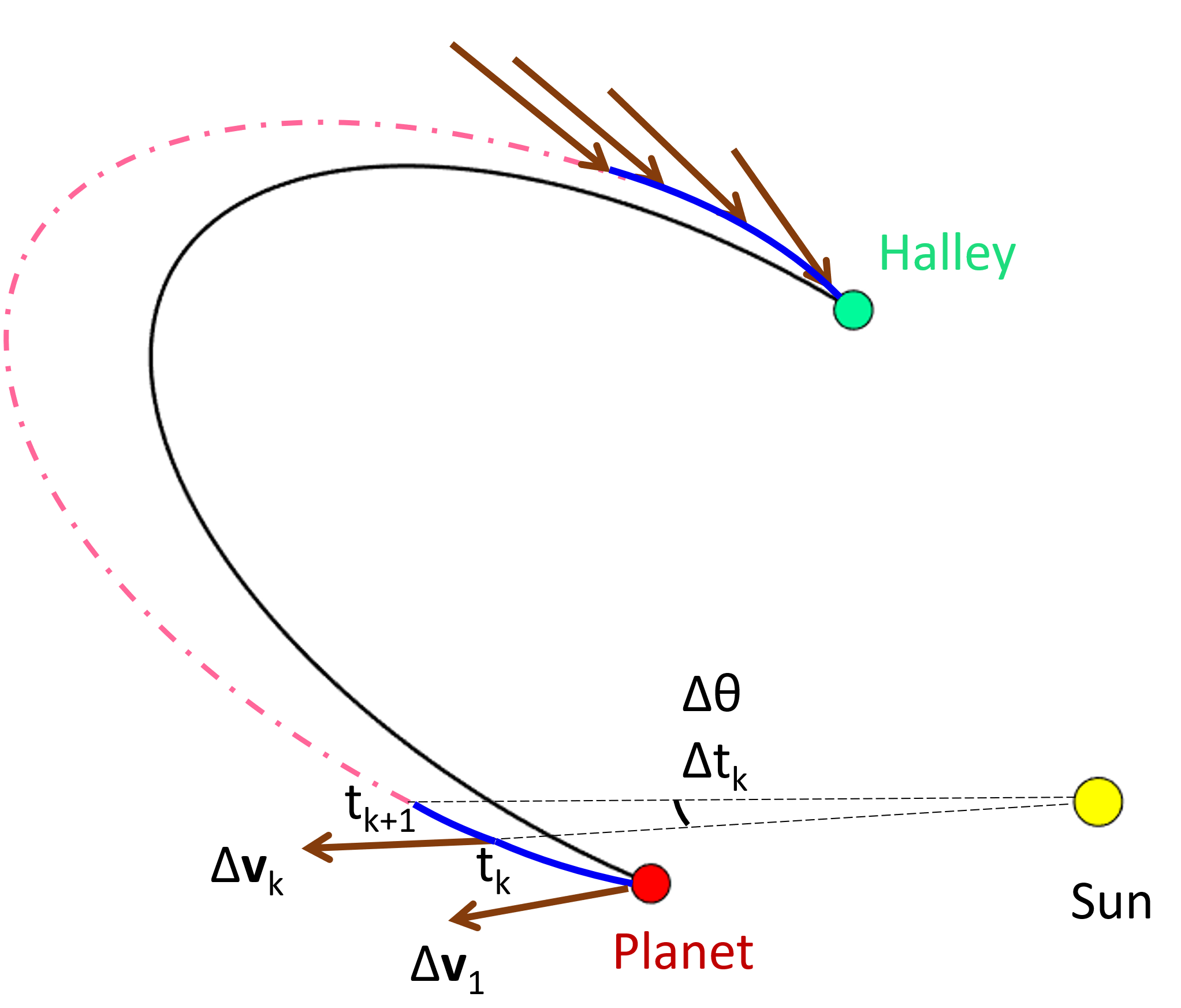}
\caption{Low-thrust design strategy: backward (left) and forward (right) propagation.}
\label{fig:DesignLT}
\end{figure}

The procedure starts from the rendezvous point, where both the spacecraft state (${\bf r}_4,{\bf v}_4$) and mass (the dry mass) are known (Fig.~\ref{fig:DesignLT} left). The direction of thrust is parameterized in terms of two angles (e.g., ecliptic longitude and latitude), and an initial guess is selected. This, together with the impulse magnitude, yields a trial value of $\Delta {\bf v}_i$, that is subtracted from the rendezvous velocity. Next, the Keplerian arc is propagated backwards in time to the previous point along the trajectory ($n-1$). The sum of the two Lambert impulses of the Keplerian arc connecting points 1 and $n-1$ ($\Delta V_{1,n-1}$ hereafter) is computed. This step is repeated for different orientations of the impulse, searching for the direction that minimizes $\Delta V_{1,n-1}$.
Next, the complete process is repeated starting from the previous point ($n-1$). That is, $\Delta V_{1,n-2}$ is minimized to determine the conditions at point $n-2$, and so on. Eventually, the algorithm will stall at point $n_s$, meaning no further reduction of the residual impulse $\Delta V_{1,n_s-1}$ is possible. This happens because it is not possible to cancel out the residual impulse completely applying thrust only at one end of the trajectory.

After the stall, the direction of propagation is switched to forward in time between points 1 (post-GA state) and $n_s$ (Fig.~\ref{fig:DesignLT} right). That is, $\Delta {\bf v}_1$ is computed minimizing $\Delta V_{2,n_s}$, this determines the state at point 2, and so on.
The advancing branch may also stall, in which case the algorithm reverts to backwards propagation.
The calculation continues until the residual impulse drops below a preset tolerance. The corresponding Lambert arc is a coast segment.

Note that the forward propagation branch requires an estimation of the wet mass in order to compute the acceleration. The first guess is obtained with the rocket equation, using $\Delta V_{\textrm{tot}}$ from the impulsive solution, the dry mass of the spacecraft and the specific impulse of the thruster (Table~\ref{tab:eng}). This initial estimate is refined iteratively, using the total impulse from the LT solutions.

\section{Results and discussion}
\label{sec_res_dis} 
This section presents two different sets of solutions. One set uses Jupiter for the GA and the other Saturn. For each flyby date, the trajectories requiring the minimum $\Delta V_{\textrm{tot}}$ have been computed. There are usually multiple solutions for each $t_2$, each corresponding to a different departure energy.

A preliminary study \cite{Beolchi:2024} showed that, for the trajectories that strike a good balance between $\Delta V_{\textrm{tot}}$ and $t_4$, the GA takes place near the major axis of the comet's orbit. Roughly speaking, the position of the planet at the time of flyby is inside, or very close to, the projection of the orbit of the comet on the ecliptic plane. To cover this range with ample margin, $t_2$ has been chosen such that the ecliptic longitude of the planet lies within $\pm 40^\circ$ of the comet's aphelion longitude.
The maximum launch energy allowed is $C3_{\max} = 250$ km$^2$/s$^2$ with a minimum GA pericenter radius $r_{\pi,\min} = 100 \, 000$ km. The search for the optimal rendezvous date is constrained to the interval between June 2053 (Uranus' orbit crossing) and May 2061 (orbit of Mars).

\subsection{Earth-Jupiter-Halley trajectories}
\label{sec_jup_hal}

\subsubsection{Impulsive trajectories}
\label{sec_jup_hal_imp} 
\begin{figure}[h!]
	\centering
	\includegraphics[width=0.94\textwidth]{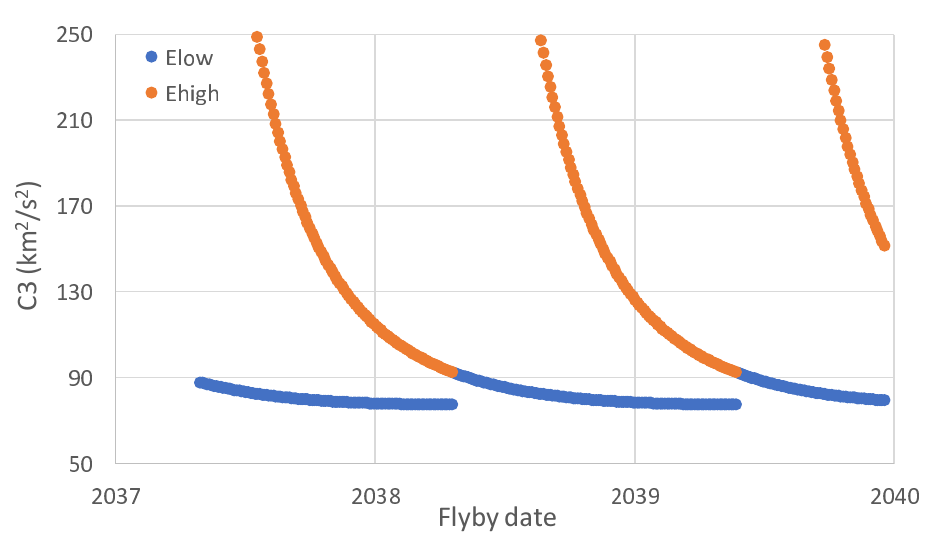}
	\caption{Earth-Jupiter transfer. Launch energy ($C3$) vs. Flyby date.}
	\label{fig_jup_t2_c3}
\end{figure}

\begin{figure}[h!]
	\centering
	\includegraphics[width=0.94\textwidth]{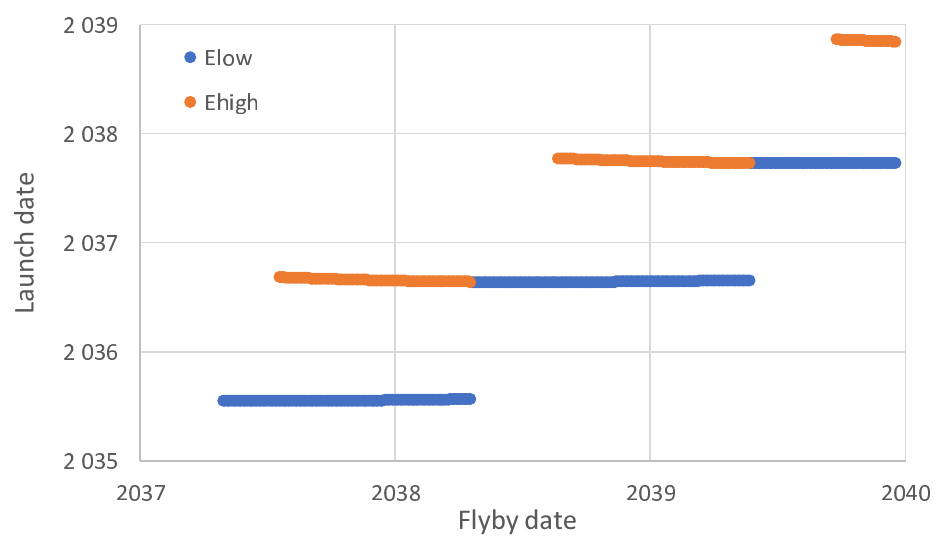}
	\caption{Earth-Jupiter transfer. Launch date vs. Flyby date.}
	\label{fig_jup_t2_t1}
\end{figure}

The range of flyby dates for Jupiter is between April 2027 and December 2039. Three hundred equally-spaced values of $t_2$ have been analyzed over this interval. There is an additional window 12 years later, but it results in flyby and rendezvous dates too close for the propulsion system to deliver enough $\Delta V$. 
Plotting the departure energy against the flyby date yields Fig.~\ref{fig_jup_t2_c3}. For each flyby date, there are two feasible values of the departure energy. They are labeled as Elow (blue markers) and Ehigh (red) in the figure. As shown in Fig.~\ref{fig_jup_t2_t1}, the two launch dates differ by 13 months approximately, which is very close to the orbital period of the Earth. The reason is that the transfer angle $\theta_{12}$ only changes from $180^\circ$ to $114^\circ$ over the range of eccentricities considered (0.677 to 1.34), while the transfer time $t_{12}$ varies between 2.7 and 0.86 years. Therefore, the left-hand-side of Eq.~\ref{eq_e} depends weakly of the eccentricity, while the longitude of the Earth changes rapidly. Thus, the position of the Earth dominates the solution, and the separation between departure dates is close to one year. It is worth noting that, due to the upper bound on $C3$, there are GA dates with no high-energy transfer. On the other hand, the low-energy solution always exists.


\begin{figure}[h!]
	\centering
	\includegraphics[width=0.94\textwidth]{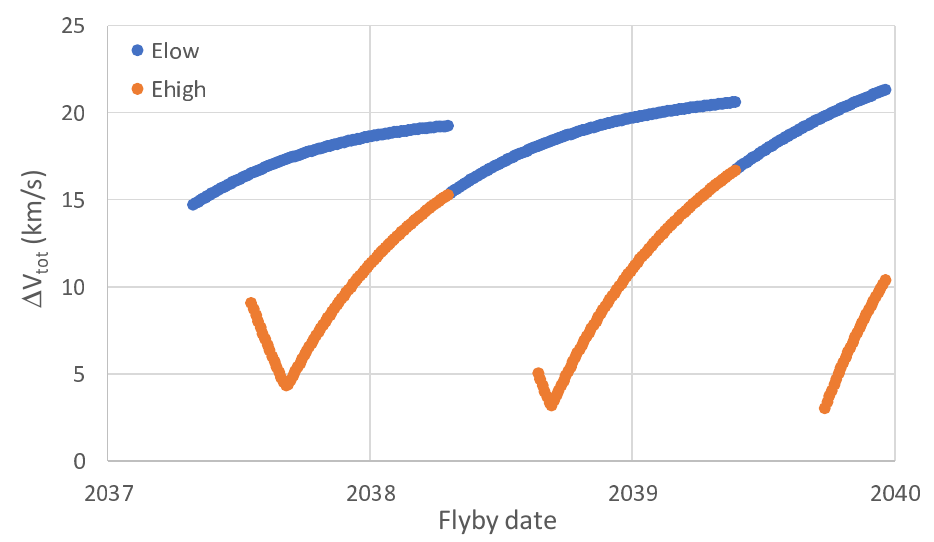}
	\caption{Earth-Jupiter-Halley trajectory. Total impulse ($\Delta V_{\textrm{tot}}$) vs. Flyby date.}
	\label{fig_jup_t2_dv}
\end{figure}

Figure~\ref{fig_jup_t2_dv} shows the variation of $\Delta V_{\textrm{tot}}$ with respect to the flyby date. The low-energy transfers between Earth and Jupiter (blue curves) lead to unfeasible trajectories, because the impulse needed vastly exceeds the capabilities of the propulsion system (5 km/s over the duration of the Jupiter-Halley transfer). The reason behind the elevated $\Delta V_{\textrm{tot}}$ is the small $v_\infty$ of the low-$C3$ transfers, which reduces the effectiveness of the GA maneuver. Hereafter, only the high-$C3$ solutions are presented to reduce clutter in the figures.

\begin{figure}[h!]
	\centering
	\includegraphics[width=0.94\textwidth]{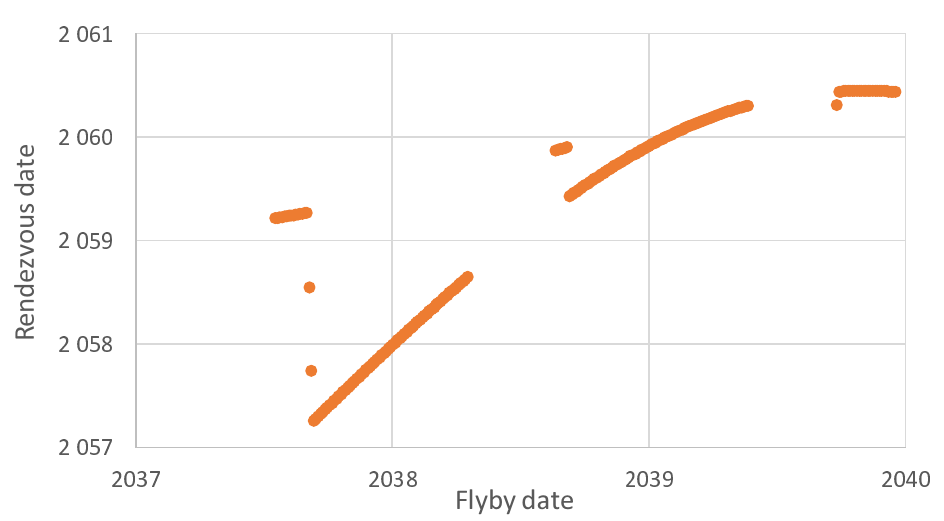}
	\caption{Earth-Jupiter-Halley trajectory. Rendezvous date vs. Flyby date.}
	\label{fig_jup_t2_t4}
\end{figure}

The rendezvous time is shown in Fig.~\ref{fig_jup_t2_t4}. To maximize the scientific return of the mission, it is desirable to reach the comet as soon as possible, allowing observation of early activity outbursts. To this effect, the flyby date should be towards the end of 2037 or beginning of 2038. Note that the right end of the curve is horizontal. This is caused by the constraint that the arrival should occur before the crossing of Mars' orbit. There is an overall trend of rendezvous date increasing progressively with flyby date, interspersed with transitions to a later arrival date near the extremes of the feasible $t_2$ ranges. These transitions shall be discussed in more detail in Sect.~\ref{sec_sat_hal_imp}, because they are easier to visualize for trajectories with Saturn GA.  

\begin{figure}[h!]
	\centering
	\includegraphics[width=0.94\textwidth]{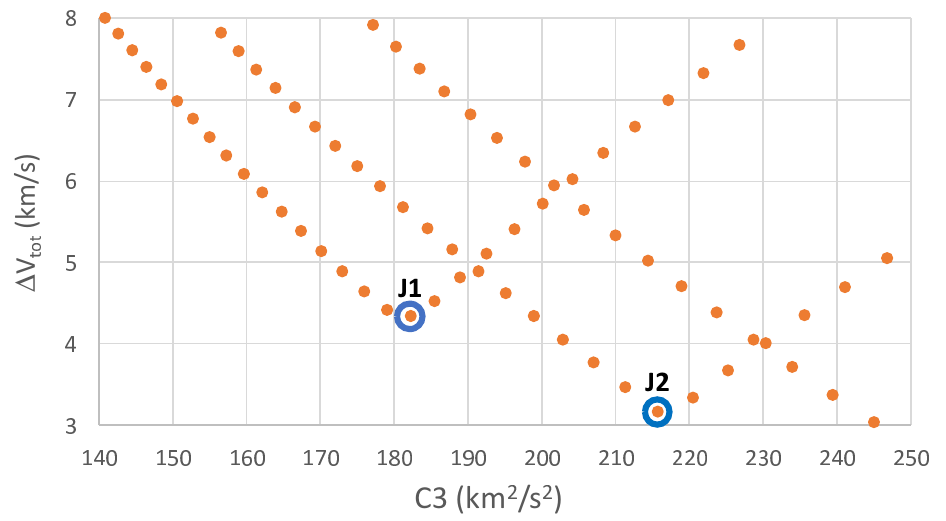}
	\caption{Earth-Jupiter-Halley trajectory. Total impulse ($\Delta V_{\textrm{tot}}$) vs. Launch energy ($C3$).}
	\label{fig_jup_c3_dv}
\end{figure}

To facilitate the selection of the most advantageous mission parameters, it is useful to plot $\Delta V_{\textrm{tot}}$ as a function of $C3$ (Fig.~\ref{fig_jup_c3_dv}). Two promising solutions have been labeled J1 and J2. J1 is characterized by moderate energy and impulse, while J2 achieves lower impulse at the cost of higher energy (see Table~\ref{tab_j12}). 

\begin{table}[h!]
	\centering
	\caption{Earth-Jupiter-Halley trajectory. Candidate solutions.}
	\begin{tabular}{|l|r|r|}
		\hline
		Solution & J1 & J2  \\ \hline
		$C3$ (km$^3$/s$^2$) & 182 & 216 \\ \hline
		$\Delta V_{\textrm{tot}}$ (km/s) & 4.34 & 3.17 \\ \hline
		$r_\pi$ ($10^5$ km) & 11.2 & 5.69 \\ \hline
		$\alpha$ (degree) & 191 & 186 \\ \hline
		Launch date & 4 Sep 2036  & 10 Oct 2037 \\ \hline
		Flyby date & 5 Sep 2037  & 11 Sep 2038 \\ \hline
		Rendezvous date & 17 Jul 2058  & 4 Jun 2059 \\ \hline
	\end{tabular}
	\label{tab_j12}
\end{table} 

Both solutions are characterized by GA inclinations close to, and above, 180$^\circ$. This means the flyby maneuver is almost leading, providing a large reduction of the spacecraft's circumferential velocity (it has to become negative in order to achieve a retrograde orbit). Furthermore, the inclination above 180$^\circ$ results in a negative z-component of the post-GA velocity, which brings the trajectory closer to the comet's orbital plane.
Given the higher efficiency of electrical propulsion compared with chemical rockets, option J1 is more attractive technically. Its higher impulse is more than compensated for by the lower $C3$, which would allow the launcher to inject a much larger mass into the Jupiter transfer arc. Furthermore, J1 rejoins the comet almost 11 months before J2, improving the scientific return. J2 has been retained mostly to illustrate an alternative trajectory. Furthermore, by virtue of its later launch date, J2 allows for an additional year of mission planning.

\subsubsection{Low-thrust trajectories}
\label{sec_jup_hal_lt} 
The impulsive trajectories have been converted into low-thrust arcs with the procedure outlined in Sect.~\ref{sec_ltt}. The switching between forward and backward propagation continues until the $\Delta V$ residual falls below 2 m/s. Because the spacecraft mass immediately after the flyby is unknown a priori, it is determined iteratively. The initial guess of the propellant mass is computed using the rocket equation and $\Delta V_{\textrm{tot}}$ from the impulsive solution. This estimate serves to compute the thrust acceleration during the forward propagation steps. Because the total impulse of the low-thrust trajectory is different from the impulsive solution, there will be a mass mismatch between the forward and backward stages. Using the secant method, the propellant mass is corrected until the difference becomes less than 0.5 kg.

\begin{figure}[h!]
	\centering
	\includegraphics[width=0.43\textwidth]{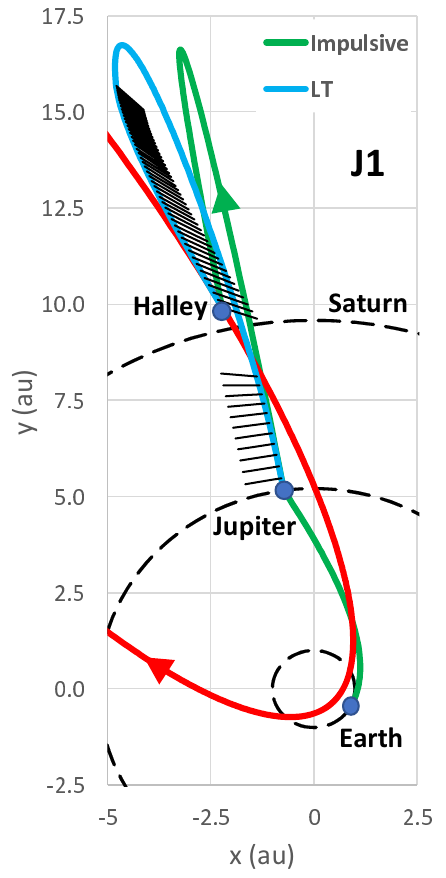}
	\includegraphics[width=0.51\textwidth]{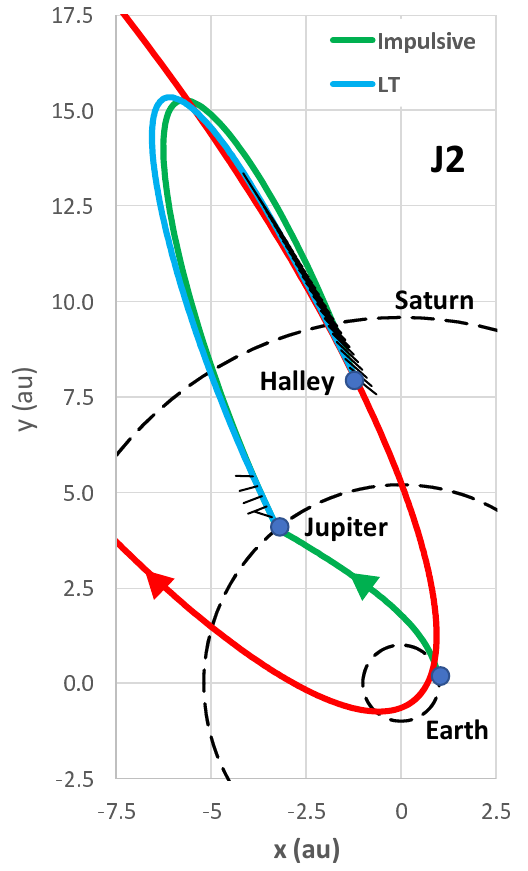}
	\caption{J1 (left) and J2 (right) trajectories. Ecliptic projection.}
	\label{fig_j12_xy}
\end{figure}

Figure \ref{fig_j12_xy} shows the ecliptic projections of the J1 and J2 trajectories. The impulsive solution is shown in green for reference while the LT trajectory is colored blue. The black lines indicate the orientation of the thrust vector, while the red curve is the orbit pf 1P/Halley.
In the case of J1, the post-GA ecliptic velocity is almost radial. The small circumferential velocity makes it possible to transition to a retrograde orbit with the electric propulsion system. The spacecraft approaches the comet's orbit from the outside, accelerating to match the comet's velocity (the spacecraft has lower specific energy than 1P/Halley after the flyby, it must increase its velocity in order to rendezvous). The rendezvous takes place ourside the orbit of Saturn, providing an excellent opportunity to study the initial stages of cometary activity.
Trajectory J2 is characterized by a larger deflection of the heliocentric velocity during the GA. In fact, the orbit is so retrograde after the encounter with Jupiter, that the thruster initially applies clockwise torque. The coast arc is almost tangent to the comet's orbit (in the ecliptic plane projection), so the rendezvous burn serves fundamentally to increase the spacecraft velocity (the thrust is aligned with the trajectory in this projection).

\begin{figure}[h!]
	\centering
	\includegraphics[width=0.94\textwidth]{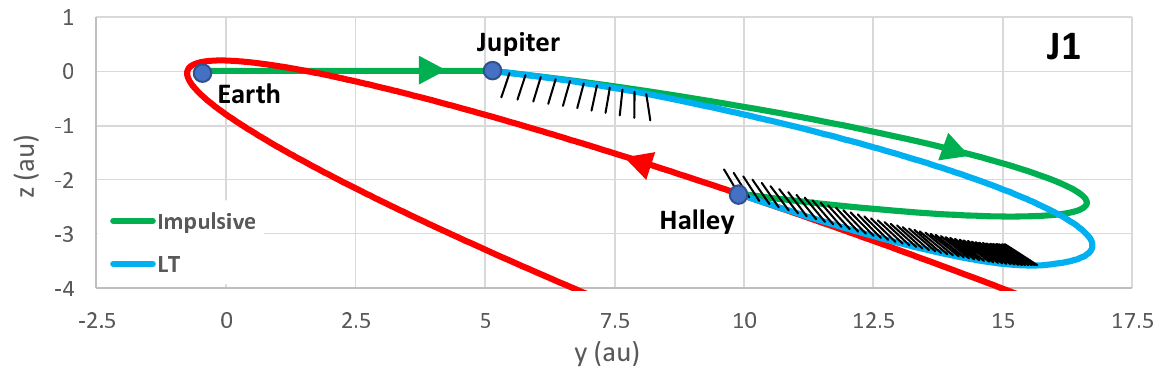}
	\includegraphics[width=0.94\textwidth]{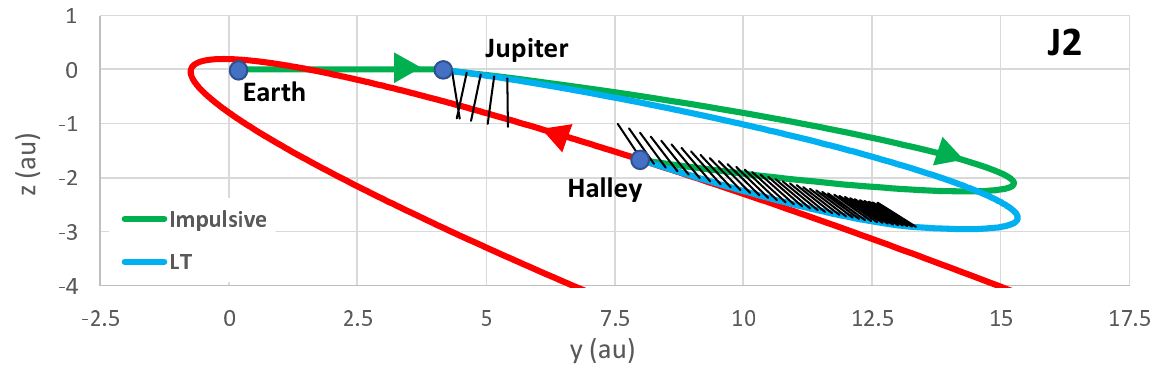}
	\caption{J1 (top) and J2 (bottom) trajectories. YZ projection.}
	\label{fig_j12_yz}
\end{figure}

Figure \ref{fig_j12_yz} depicts the YZ projections of the trajectories, which have a similar arrangement in this view. The flyby curves the trajectory downwards to bring it closer to 1P/Halley, which approaches the Sun from below the ecliptic. Immediately after the GA, the thrust vector is directed towards the South of the ecliptic, to increase the downward velocity even further. The final burn requires upwards thrust to mach the ascending velocity of the comet.

\begin{table}[h!]
	\centering
	\caption{Performance of Earth-Jupiter-Halley impulsive and LT solutions.}
	\begin{tabular}{|l|r|r|}
		\hline
		Solution & J1 & J2  \\ \hline
		Impulsive $\Delta V_{\textrm{tot}}$ (km/s) & 4.34 & 3.17 \\ \hline
		LT $\Delta V_{\textrm{tot}}$ (km/s) & 5.79 & 4.13 \\ \hline
		LT/Imp. $\Delta V_{\textrm{tot}}$ ratio & 1.33 & 1.31 \\ \hline
		Propellant mass (kg) & 446 & 301 \\ \hline
	\end{tabular}
	\label{tab_jup_ivlt}
\end{table} 

Table \ref{tab_jup_ivlt} compares the performance of the LT and impulsive solutions. The continuous thrust trajectories always require more impulse (over 30\% larger in this case), but this is completely offset by the vastly superior specific impulse of electric thrusters. This keeps the propellant budget within feasible limits (45\% and 30\% of the dry mass for J1 and J2, respectively). The efficiency of the LT arcs becomes closer to the impulsive maneuver as $\Delta V_{\textrm{tot}}$ decreases (in fact, they are identical in the limit of infinitesimally small impulse). This will be further illustrated by the trajectories with Saturn GA, which a characterized by smaller total impulses.

\subsection{Earth-Saturn-Halley trajectories}
\label{sec_sat_hal} 
\begin{figure}[h!]
	\centering
	\includegraphics[width=0.94\textwidth]{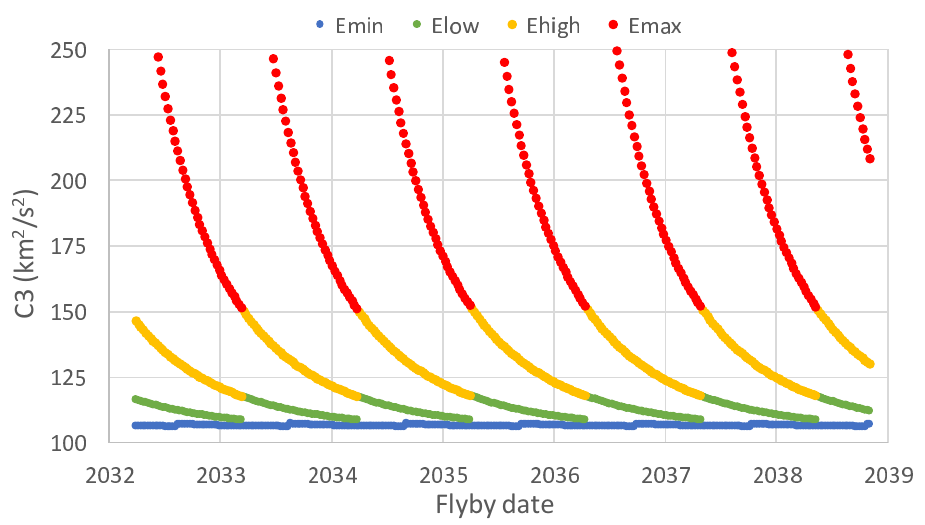}
	\caption{Earth-Saturn transfer. Launch energy ($C3$) vs. Flyby date.}
	\label{fig_sat_t2_c3}
\end{figure}

In the case of Saturn, the interval of GA dates is from March 2032 to November 2038. This is much wider than for Jupiter due to the slower mean motion of Saturn. The $C3-t_2$ plot is shown in Fig.~\ref{fig_sat_t2_c3}. In this case, there are between 3 and 4 $C3$ levels for each GA date. They are denoted as Emin/Elow/Ehigh/Emax in the figure. As was the case before, some flyby dates are not possible with the hig-energy solution due to the limit placed on $C3$.

\subsubsection{Impulsive trajectories}
\label{sec_sat_hal_imp} 
\begin{figure}[h!]
	\centering
	\includegraphics[width=0.94\textwidth]{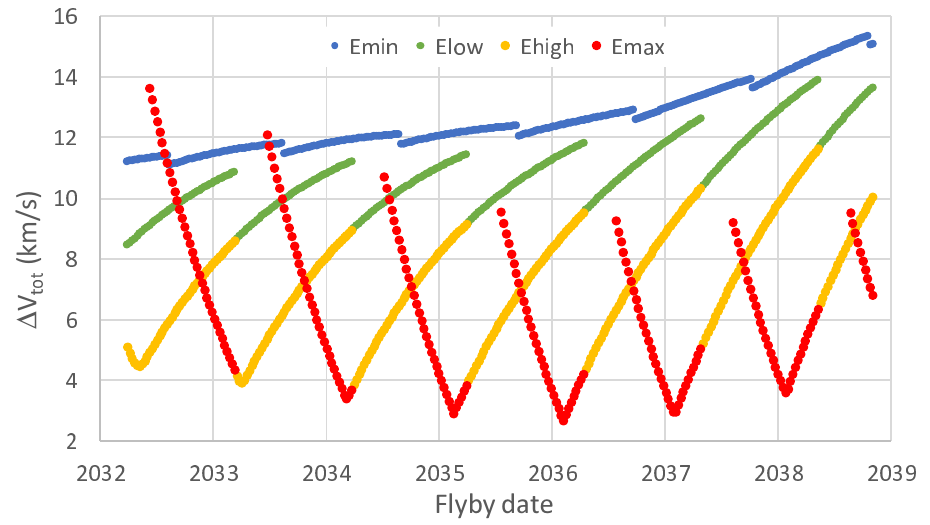}
	\caption{Earth-Saturn-Halley trajectory. Total impulse ($\Delta V_{\textrm{tot}}$) vs. Flyby date.}
	\label{fig_sat_t2_dv}
\end{figure}

As seen in Fig.~\ref{fig_sat_t2_dv}, only the two highest energies yield acceptable values of $\Delta V_{\textrm{tot}}$. The two solution with the lowest $C3$ will be removed from this point on, for the sake of clarity. The Saturn GA is advantageous from the point of view of $\Delta V_{\textrm{tot}}$. A value of 2.65 km/s is possible, compared to 3.17 km/s for Jupiter.

\begin{figure}[h!]
	\centering
	\includegraphics[width=0.94\textwidth]{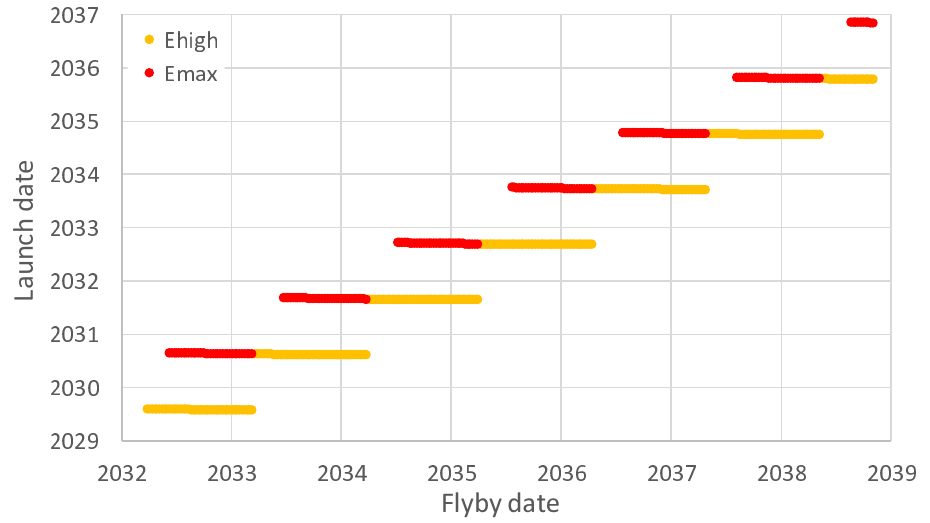}
	\caption{Earth-Saturn transfer. Launch date vs. Flyby date.}
	\label{fig_sat_t2_t1}
\end{figure}

As in the case of Jupiter, the feasible launch dates for each $t_2$ differ by slightly more than one year (Fig.~\ref{fig_sat_t2_t1}). Unfortunately, Saturn trajectories require launching earlier than in the case of Jupiter, leaving less time to prepare the mission.

\begin{figure}[h!]
	\centering
	\includegraphics[width=0.94\textwidth]{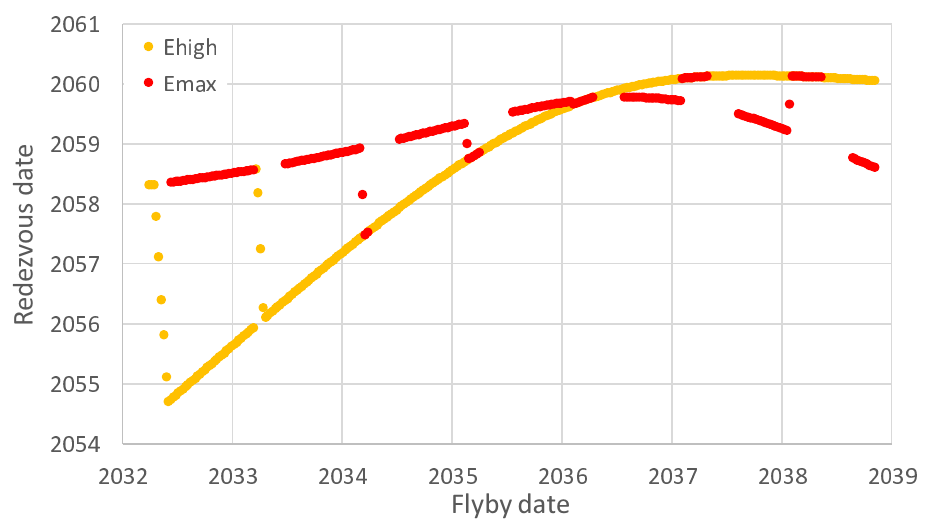}
	\caption{Earth-Saturn-Halley trajectory. Rendezvous date vs. Flyby date.}
	\label{fig_sat_t2_t4}
\end{figure}

The rendezvous time curve (Fig.~\ref{fig_sat_t2_t4}) shows two distinct branches, with sharp transitions between them. The sudden changes occur at the local minima of $\Delta V_{\textrm{tot}}$ in Fig.~\ref{fig_sat_t2_dv} (the cusps in the curves stem from the switch of regime). Note that, for some values of $t_2$, both solutions (Ehigh and Emax) have virtually the same $t_4$. This happens because the apex of the $\Delta V_{\textrm{tot}}$ curve does not coincide with the boundary between the red and orange curves. Thus, the two solutions can be in the same regime.

\begin{figure}[h!]
	\centering
	\includegraphics[width=0.94\textwidth]{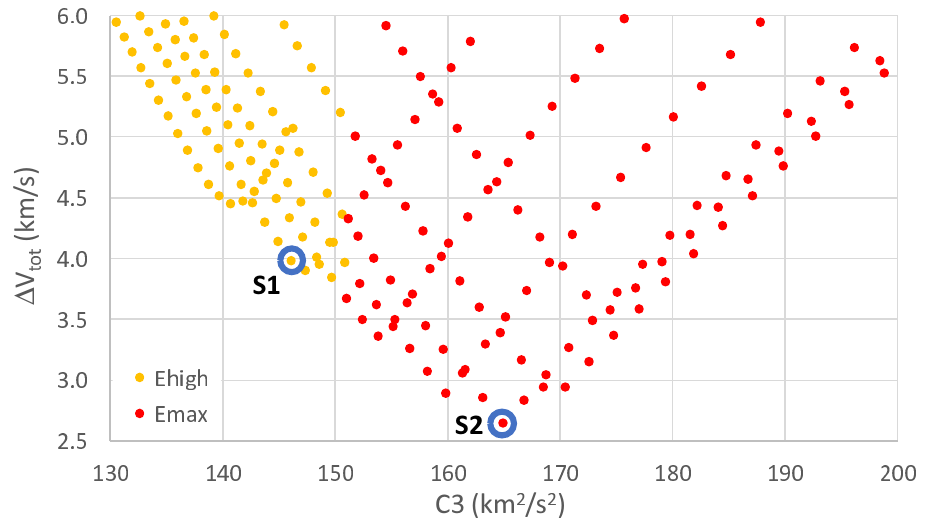}
	\caption{Earth-Saturn-Halley trajectory. Total impulse ($\Delta V_{\textrm{tot}}$) vs. Launch energy ($C3$).}
	\label{fig_sat_c3_dv}
\end{figure}

Finally, Fig.~\ref{fig_sat_c3_dv} combines $\Delta V_{\textrm{tot}}$ and $C3$ to assist in the selection of the trajectory offering the best trade-off between the different parameters. Two candidates delivering low $C3$ and $\Delta V_{\textrm{tot}}$ have been marked as S1 and S2 (see Table~\ref{tab_s12}).

\begin{table}[h!]
	\centering
	\caption{Earth-Saturn-Halley trajectory. Candidate solutions.}
	\begin{tabular}{|l|r|r|}
		\hline
		Solution & S1 & S2  \\ \hline
		$C3$ (km$^3$/s$^2$) & 146 & 165 \\ \hline
		$\Delta V_{\textrm{tot}}$ (km/s) & 3.98 & 2.65 \\ \hline
		$r_\pi$ ($10^5$ km) & 10.8 & 3.89 \\ \hline
		$\alpha$ (degree) & 199 & 186 \\ \hline
		Launch date & 17 Aug 2030  & 25 Sep 2033 \\ \hline
		Flyby date & 13 Apr 2033  & 8 Feb 2036 \\ \hline
		Rendezvous date & 5 Apr 2056  & 26 Aug 2059 \\ \hline
	\end{tabular}
	\label{tab_s12}
\end{table} 

Note that S1 is not the local minimum of $\Delta V_{\textrm{tot}}$, but the point immediately to the left of it. The transition between the two branches of the rendezvous time curve in Fig.~\ref{fig_sat_t2_t4} takes place at the minimum. Moving slightly away from it pulls back $t_4$ by a full year at practically no cost ($\Delta V_{\textrm{tot}}$ increases less than 0.1 km/s, while $C3$ decreases slightly). Thus, S1 is more attractive than the actual minimum.
S2 delivers the global minimum of impulse, but with a later arrival at the comet. In this case, moving away from S1 does not bring substantial gains because the minimum of $\Delta V_{\textrm{tot}}$ coincides with the intersection of the two branches of the $t_4$ curve (see Fig.~\ref{fig_sat_t2_t4}). Thus, switching branches has a very limited effect on the rendezvous date.

\subsubsection{Low-thrust trajectories}
\label{sec_sat_hal_lt} 
\begin{figure}[h!]
	\centering
	\includegraphics[width=0.43\textwidth]{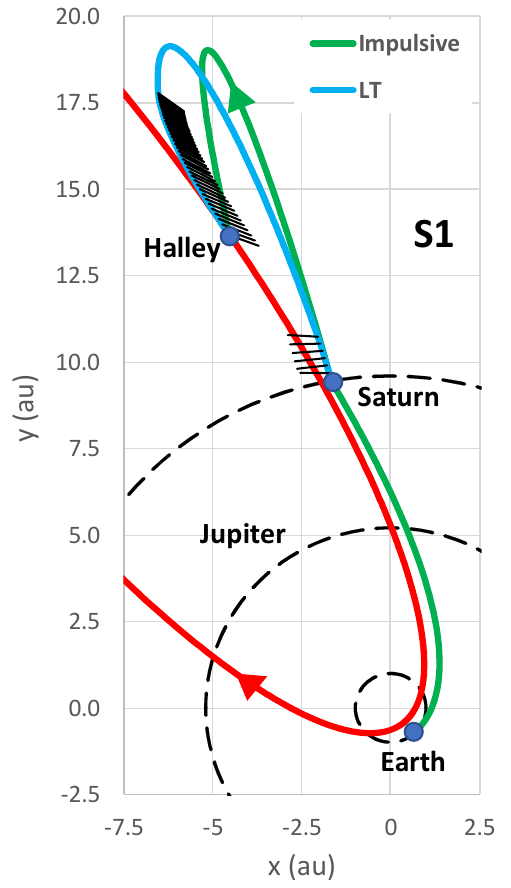}
	\includegraphics[width=0.51\textwidth]{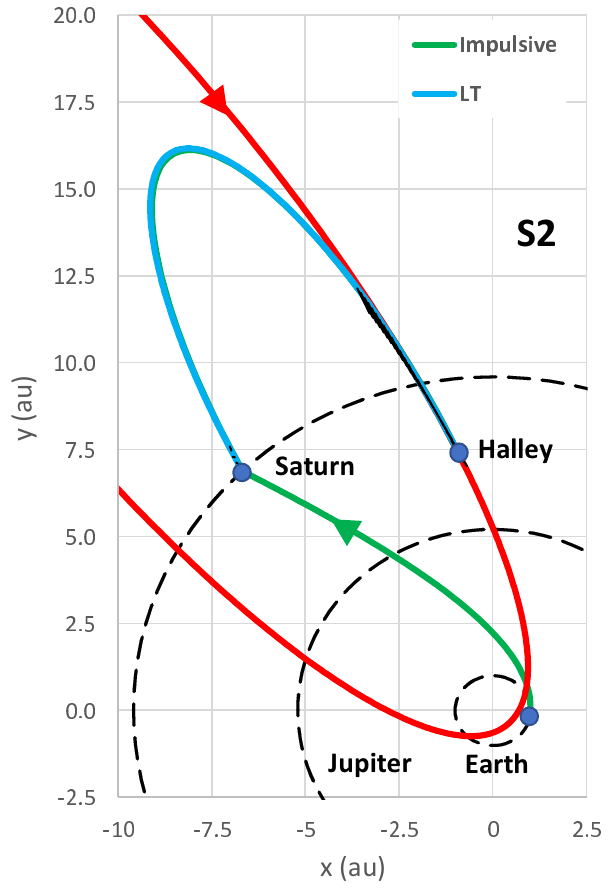}
	\caption{S1 (left) and S2 (right) trajectories. Ecliptic projection.}
	\label{fig_s12_xy}
\end{figure}

The ecliptic projections of the S1 and S2 solutions are shown in Fig.~\ref*{fig_s12_xy}. They follow a pattern similar to the Jupiter flyby trajectories. S1 leaves Saturn in an almost radial direction, the trajectory at this point is still prograde, but the small circumferential velocity makes it easy to transition to retrograde motion. Then, the probe reaches the comet from the outside of its orbit. The Saturn solutions are advantageous because, due to the smaller heliocentric velocities when the maneouvers are applied far from the Sun, it is less expensive to perform the plane change (both in terms of $C3$ and $\Delta V_{\textrm{tot}}$). Furthermore, S1 encounters 1P/Halley 14.8 au away from the Sun, increasing the chance of observing unexpected early bursts of activity.
S2 follows a strategy comparable to J2, with a strong change in direction during the GA that places the spacecraft directly in a retrograde trajectory tangent to the comet's orbit, approaching it from the inside. Interestingly, the ecliptic projections of the impulsive and LT solutions are almost indistinguishable. This means the in-plane component of the first Lambert impulse (${\bf v}^L_3 - {\bf v}_3$) is very small. That is, the GA is able to match the in-plane component of ${\bf v}^L_3$ with high accuracy. This is enabled by the flyby maneuver far from the Sun, that reduces the magnitude of the corrections. As shown in the figure, almost all the in-plane thrust is spent increasing the spacecraft velocity during the rendezvous burn. 

\begin{figure}[h!]
	\centering
	\includegraphics[width=0.94\textwidth]{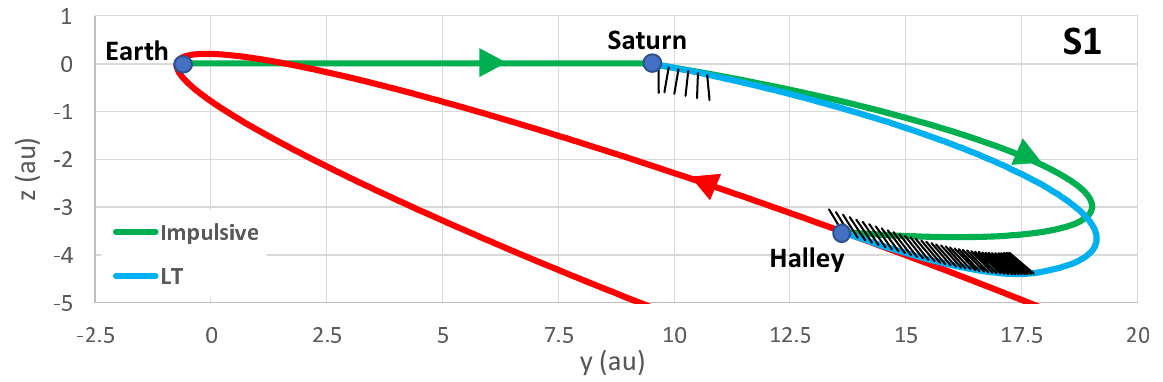}
	\includegraphics[width=0.94\textwidth]{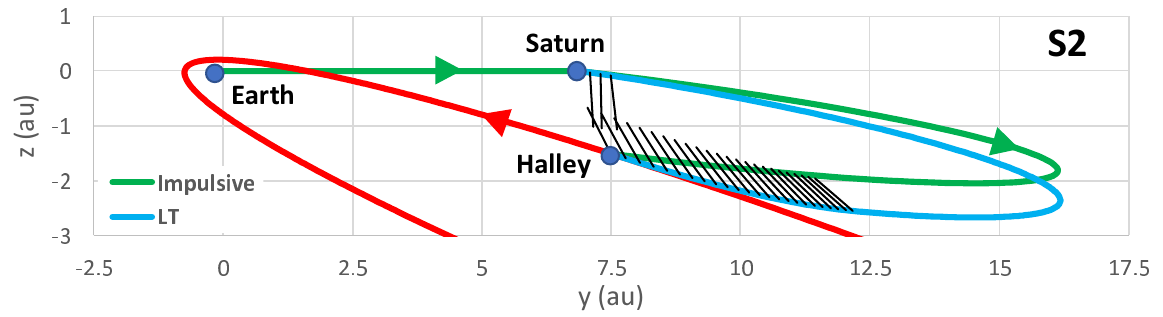}
	\caption{S1 (top) and S2 (bottom) trajectories. YZ projection.}
	\label{fig_s12_yz}
\end{figure}

The YZ projections can be found in Fig.~\ref*{fig_s12_yz}. The behavior is again similar to J1 and J2, with the trajectories first curving downwards to approach the comet, and then upwards to match its ascending velocity.

\begin{table}[h!]
	\centering
	\caption{Performance of Earth-Saturn-Halley impulsive and LT solutions.}
	\begin{tabular}{|l|r|r|}
		\hline
		Solution & S1 & S2  \\ \hline
		Impulsive $\Delta V_{\textrm{tot}}$ (km/s) & 3.98 & 2.65 \\ \hline
		LT $\Delta V_{\textrm{tot}}$ (km/s) & 4.95 & 3.15 \\ \hline
		LT/Imp. $\Delta V_{\textrm{tot}}$ ratio & 1.24 & 1.19 \\ \hline
		Propellant mass (kg) & 371 & 222 \\ \hline
	\end{tabular}
	\label{tab_sat_ivlt}
\end{table}

Table \ref*{tab_sat_ivlt} shows the performance of the LT trajectories. As indicated before, the lower impulse of the Saturn-flyby solutions make the transition to continuous thrust more efficient, requiring only 19\% to 24\% more $\Delta V_{\textrm{tot}}$ than the high-thrust maneuver.

\section{Conclusions}
\label{sec_con} 
We presented a strategy for a rendezvous mission with comet Halley that would allow close observation of the start of activity prior to the next perihelion (year 2061). Contrary to other studies, which assume hardware currently unavailable (such as high-power nuclear reactors), the mission planning relies on existing Hall-effect thrusters powered by standard RTGs (meaning the power budget is limited).

The design strategy combines a flyby with Jupiter or Saturn with continuous thust arcs to reach the comet with zero relative velocity before it crosses the orbit of Mars (when the high-activity stage is expected to begin). A dry spacecraft mass of 1000 kg is assumed, together with a maximum thrust of 36 mN and specific impulse of 1600 s. The trajectory is first optimized assuming impulsive maneuvers. Taking the flyby date as a free design parameter, it is possible to determine the optimal gravity-assist geometry as a function of the rendezvous time only. Thus, the optimization process needs to adjust only one parameter and becomes very efficient. The optimal impulsive trajectory is then converted into continuous thrust arcs using a strategy that propagates the trajectory backwards and forward in time, searching for the thrust direction that achieves the fastest reduction of the residual impulse. 

Four candidate trajectories that achieve the mission goals with reasonable values of the launch characteristic energy ($C3$) and impulse of the low-thrust maneuvers ($\Delta V_{\textrm{tot}}$) have been presented.
The $C3$ requirement ranges from 146 to 215 km$^2$/s$^2$, depending on the trajectory configuration (flyby planet and date) with a wet mass below 1500 kg. While $C3$ is high, it is well within the capabilities of modern heavy launchers. For example, the SLS Block 2 with a Centaur kick stage can inject a mass of 2500 kg at 215 km$^2$/s$^2$, and over 6000 kg at 150 km$^2$/s$^2$ \cite{Stough:2021}. In fact, the lower bound of $C3$ is not unheard of, as New Horizons achieved 157 km$^2$/s$^2$ \cite{Stough:2021} (albeit whith a smaller mass).
The range of impulses for the post-flyby leg is from 2.65 to 4.34 km/s for impulsive maneuvers, and 3.16 to 5.8 km/s for low thrust. This requires a propellant mass between 222 and 446 kg.

The solutions with a Saturn flyby are generally more efficient in terms of $C3$ and $\Delta V_{\textrm{tot}}$, due to the smaller heliocentric velocities farther from the Sun. They can also reach the comet earlier but, unfortunately, allow for shorter mission planning time (earlier launch date than with a Jupiter flyby). 

Probably, the most promising mission profile of those discussed is S1. It requires a very reasonable $C3$ of 146 km$^2$/s$^2$ with 371 kg of propellant, reaching the comet in early 2056, when it is almost 15 au from the Sun. Therefore, it offers great scientific potential, enabling observation of the earliest stages of cometary activity. Unfortunately, the launch date is near (mid-2030), so the window of opportunity is closing fast.
The Jupiter flyby, while energetically less attractive, enables more time for mission planning. A launch in late 2037 is possible with $C3 = 216$ km$^2$/s$^2$ and 301 kg of propellant. It could serve as a fallback plan, if launching in the early 2030s is not an option.

This work serves as a proof of concept, demonstrating that this kind of mission is possible with existing technology, provided the project is launched very soon.

\section*{Acknowledgments}
R. Flores, A. Beolchi, E. Fantino and C. Pozzi acknowledge Khalifa University of Science and Technology's internal grant CIRA-2021-65/8474000413. A. Beolchi and E. Fantino have received support from project 8434000368 (Khalifa University of Science and Technology's 6U Cubesat mission). R. Flores  and  E. Fantino acknowledge grant ELLIPSE/8434000533 (Abu Dhabi's Technology Innovation Institute). E. Fantino has been partially supported by project PID2021-123968NB-I00 (Spanish Ministry of Science and Innovation). The work of C. Barbieri has been partially funded by a visiting scholar contract at Khalifa University of Science and Technology under grant CIRA-2021-65/8474000413.  

\bibliography{References}

\begin{thebibliography}{10}
\expandafter\ifx\csname url\endcsname\relax
  \def\url#1{\texttt{#1}}\fi
\expandafter\ifx\csname urlprefix\endcsname\relax\def\urlprefix{URL }\fi
\expandafter\ifx\csname href\endcsname\relax
  \def\href#1#2{#2} \def\path#1{#1}\fi

\bibitem{Barbieri:2017}
C.~Barbieri, I.~Bertini, Comets, La Rivista del Nuovo Cimento 40 (2017)
  335--409.
\newblock \href {https://doi.org/10.1393/ncr/i2017-10138-4}
  {\path{doi:10.1393/ncr/i2017-10138-4}}.

\bibitem{Brady:1971}
J.~L. Brady, E.~Carpenter, {The Orbit of Halley's Comet and the Apparition of
  1986}, Astronomical Journal 76 (1971) 728--739.
\newblock \href {https://doi.org/10.1086/111190} {\path{doi:10.1086/111190}}.

\bibitem{Yeomans:1981}
D.~K. Yeomans, T.~Kiang, {The long-term motion of comet Halley}, Monthly
  Notices of the Royal Astronomical Society 197~(3) (1981) 633--646.
\newblock \href {https://doi.org/10.1093/mnras/197.3.633}
  {\path{doi:10.1093/mnras/197.3.633}}.

\bibitem{Reinhard:1982}
R.~Reinhard, {The Giotto mission to Halley's comet}, Advances in Space Research
  2~(12) (1982) 97--107.
\newblock \href {https://doi.org/10.1016/0273-1177(82)90293-9}
  {\path{doi:10.1016/0273-1177(82)90293-9}}.

\bibitem{Stelzried:1986}
C.~{Stelzried}, L.~{Efron}, J.~{Ellis}, {Halley Comet Missions},
  Telecommunications and Data Acquisition Progress Report 87 (1986) 240--242.

\bibitem{Oya:1986}
H.~Oya, {Japanese Halley missions with Sakigake and Suisei}, Eos, Transactions
  American Geophysical Union 67~(6) (1986) 65--66.
\newblock \href {https://doi.org/10.1029/EO067i006p00065-01}
  {\path{doi:10.1029/EO067i006p00065-01}}.

\bibitem{Sagdeev:1986}
R.~Sagdeev, J.~Blamont, A.~Galeev, V.~Kovtunenko, V.~Moroz, V.~Shapiro,
  V.~Shevchenko, K.~Szego, {VEGA-1 and VEGA-2 Spacecraft Encounters with Comet
  Halley}, Soviet Astronomy Letters 12 (1986) 243--247.

\bibitem{Keller:1986}
H.~U. Keller, C.~Arpigny, C.~Barbieri, R.~Bonnet, S.~Cazes, M.~Coradini,
  C.~Cosmovici, W.~Delamere, W.~Huebner, D.~Hughes, et~al., {First Halley
  multicolour camera imaging results from Giotto}, Nature 321 (1986) 320--326.
\newblock \href {https://doi.org/10.1038/321320a0}
  {\path{doi:10.1038/321320a0}}.

\bibitem{Reinhard:1987}
R.~Reinhard, {The Giotto mission to comet Halley}, Journal of Physics E:
  Scientific Instruments 20~(6) (1987) 700--712.
\newblock \href {https://doi.org/10.1088/0022-3735/20/6/029}
  {\path{doi:10.1088/0022-3735/20/6/029}}.

\bibitem{Akiba:1980}
R.~Akiba, F.~Liann, H.~Matsuo, K.~Uesugi, M.~Ichikawa, {Orbital design and
  technological feasibility of Halley mission}, Acta Astronautica 7 (1980)
  979--805.
\newblock \href {https://doi.org/10.1016/0094-5765(80)90108-3}
  {\path{doi:10.1016/0094-5765(80)90108-3}}.

\bibitem{Reinhard:1986}
R.~Reinhard, {The Giotto encounter with comet Halley}, Nature 321 (1986)
  313--318.
\newblock \href {https://doi.org/10.1038/321313a0}
  {\path{doi:10.1038/321313a0}}.

\bibitem{Curdt:1988}
W.~Curdt, K.~Wilhelm, A.~Craubner, E.~Krahn, H.~Keller, {Position of comet
  P/Halley at the Giotto encounter}, Astronomy and Astrophysics 191 (1988)
  L1--L3.

\bibitem{Volwerk:2014}
M.~Volwerk, K.-H. Glassmeier, M.~Delva, D.~Schmid, C.~Koenders, I.~Richter,
  K.~Szeg{\"o}, {A comparison between VEGA 1, 2 and Giotto flybys of comet
  1P/Halley: implications for Rosetta}, Annales Geophysicae 32~(11) (2014)
  1441--1453.
\newblock \href {https://doi.org/10.5194/angeo-32-1441-2014}
  {\path{doi:10.5194/angeo-32-1441-2014}}.

\bibitem{Keller:2020}
H.~U. Keller, E.~K{\"u}hrt, {Cometary nuclei—from Giotto to Rosetta}, Space
  Science Reviews 216 (2020) 1--26.
\newblock \href {https://doi.org/10.1007/s11214-020-0634-6}
  {\path{doi:10.1007/s11214-020-0634-6}}.

\bibitem{Barbieri:2025}
C.~Barbieri, A.~Beolchi, I.~Bertini, V.~Da~Deppo, E.~Fantino, R.~Flores,
  C.~Pernechele, C.~Pozzi, {Preparing for the 2061 return of Halley’s
  comet---A rendezvous mission with an innovative imaging system}, {Submitted
  to Planetary and Space Science, Preprint available at arxiv.org} (2025).
\newblock \href {https://doi.org/10.48550/arXiv.2502.12816}
  {\path{doi:10.48550/arXiv.2502.12816}}.

\bibitem{Friedlander:1971}
A.~L. Friedlander, J.~C. Niehoff, J.~I. Waters, Trajectory requirements for
  comet rendezvous, Journal of Spacecraft and Rockets 8~(8) (1971) 858--866.
\newblock \href {https://doi.org/10.2514/3.59735} {\path{doi:10.2514/3.59735}}.

\bibitem{Horsewood:2023}
J.~Horsewood, R.~McNutt, A.~Delamere, {Whipple Come Halley 2061 Rendezvous
  Mission}, {NASA Technology Showcase, Galveston, Texas} (2023).

\bibitem{Kruse:1969}
D.~H. Kruse, M.~K. Fox, {Trajectory Analysis Aspects of Low-Thrust and
  Ballistic Rendezvous Missions to Halley's Comet}, in: Proceedings of the
  AIAA/AAS Astrodynamics Conference, Princeton, New Jersey, August 20-22,
  American Institute of Aeronautics and Astronautics, Paper 69-933, 1969.

\bibitem{Friedlander:1970}
A.~Friedlander, J.~Niehoff, J.~Waters, Trajectory and propulsion
  characteristics of comet rendezvous opportunities, Tech. Rep. T-25, IIT
  Research Institute, Chicago, Illinois (1970).

\bibitem{Friedlander:1971b}
A.~Friedlander, W.~Wells, Comet rendezvous mission study, Tech. Rep. M-28, IIT
  Research Institute, Chicago, Illinois (1971).

\bibitem{Michielsen:1968}
H.~Michielsen, {A rendezvous with Halley's comet in 1985-1986}, Journal of
  Spacecraft and Rockets 5~(3) (1968) 328--334.
\newblock \href {https://doi.org/10.2514/3.29247} {\path{doi:10.2514/3.29247}}.

\bibitem{Friedlander:1967}
A.~Friedlander, On the problem of comet orbit determination for spacecraft
  intercept missions, Tech. Rep. T-19, IIT Research Institute, Chicago,
  Illinois (1967).

\bibitem{Deerwester:1969}
J.~M. Deerwester, {Jupiter swingby missions to nonspecific locations in
  interplanetary space}, Tech. Rep. NASA TN D-5271, NASA Headquarters, Moffett
  Field, California (1969).

\bibitem{Manning:1970}
L.~Manning, {Comet rendezvous missions}, in: Proceedings of the AAS/AIAA
  Astrodynamics Conference, Santa Barbara, California, August 19-21, 1970,
  {Paper AIAA 70-1074}.

\bibitem{Farquhar:1977}
R.~W. Farquhar, W.~H. Wooden, et~al., {Opportunities for ballistic missions to
  Halley's comet}, Tech. Rep. Technical Memorandum 71289, Goddard Space Flight
  Center, Greenbelt, Maryland (1977).

\bibitem{Farquhar:1978}
R.~W. Farquhar, W.~H. Wooden, et~al., {Cometary exploration in the Shuttle
  era}, Tech. Rep. Technical Memorandum 78033, Goddard Space Flight Center,
  Greenbelt, Maryland (1978).

\bibitem{Jacobson:1978}
R.~A. Jacobson, C.~L. Thornton, {Elements of solar sail navigation with
  application to a Halley's comet rendezvous}, Journal of Guidance and Control
  1~(5) (1978) 365--371.
\newblock \href {https://doi.org/10.2514/3.55794} {\path{doi:10.2514/3.55794}}.

\bibitem{West:1978}
J.~West, {Ion drive technology readiness for the 1985 Halley Comet rendezvous
  mission}, in: Proceedings of the $13^{\rm th}$ International Electric
  Propulsion Conference, American Institute of Aeronautics and Astronautics,
  1978, {AIAA Paper 78-641}.
\newblock \href {https://doi.org/10.2514/6.1978-641}
  {\path{doi:10.2514/6.1978-641}}.

\bibitem{Farquhar:1979}
R.~W. Farquhar, W.~H. Wooden, {Halley's Comet 1985-86: A Unique Opportunity for
  Space Explorartion}, Tech. Rep. Technical Memorandum 80633, Goddard Space
  Flight Center, Greenbelt, Maryland (1979).

\bibitem{Farquhar:1982}
R.~Farquhar, D.~Dunham, {Earth-return trajectory options for the 1985-86 Halley
  opportunity}, in: Proceedings of the 20$^{\rm th}$ Aerospace Sciences
  Meeting, American Institute of Aeronautics and Astronautics, Orlando,
  Florida, January 11-14, 1982.
\newblock \href {https://doi.org/10.2514/6.1982-136}
  {\path{doi:10.2514/6.1982-136}}.

\bibitem{Ozaki:2020}
N.~Ozaki, et~al., {Concept Study of Comet Halley Revisiting Missions},
  {Symposium on Planetary Sciences, Tohoku University, Japan} (2020).

\bibitem{Beolchi:2024}
A.~Beolchi, C.~Pozzi, E.~Fantino, R.~Flores, C.~Barbieri, I.~Bertini,
  {Low-Thrust Gravity-Assisted Rendezvous 2061 Trajectory To Halley's Comet},
  in: Proceedings of the $75^{\rm th}$ International Astronautical Congress,
  Milan, Italy, October 14-18, International Astronautical Federation, 2024,
  {Paper IAC24-C1.8.7}.

\bibitem{horsys}
{NASA Jet Propulsion Laboratory}, {Horizons System},
  \url{https://ssd.jpl.nasa.gov/horizons/} (last accessed: 9 Sept. 2024).

\bibitem{spicenaif}
{NASA Jet Propulsion Laboratory, Navigation and Ancillary Information Facility
  (NAIF)}, {Spacecraft Planet Instrument C-matrix Events (SPICE) toolkit
  (version N0067)}, \url{https://naif.jpl.nasa.gov/naif/toolkit.html} (last
  accessed: 9 Sept. 2024) (2022).

\bibitem{Glassmeier:2007}
K.-H. Glassmeier, H.~Boehnhardt, D.~Koschny, E.~K{\"u}hrt, I.~Richter, The
  rosetta mission: flying towards the origin of the solar system, Space Science
  Reviews 128 (2007) 1--21.
\newblock \href {https://doi.org/10.1007/s11214-006-9140-8}
  {\path{doi:10.1007/s11214-006-9140-8}}.

\bibitem{Blume:2005}
W.~H. Blume, {Deep Impact Mission Design}, Space Science Reviews 117 (2005)
  23--42.
\newblock \href {https://doi.org/10.1007/s11214-005-3386-4}
  {\path{doi:10.1007/s11214-005-3386-4}}.

\bibitem{Vaudolon:2019}
J.~Vaudolon, V.~Vial, N.~Cornu, I.~Habbassi, {PPS\textcircled{R} X00 Thruster
  Development Status at Safran}, in: Proceedings of the $36^{\rm th}$
  International Electric Propulsion Conference, Vienna, Austria, 2019.

\bibitem{Fantino:2023}
E.~Fantino, B.~M. Burhani, R.~Flores, E.~M. Alessi, F.~Solano,
  M.~Sanjurjo-Rivo, {End-to-end trajectory concept for close exploration of
  Saturn's Inner Large Moons}, Communications in Nonlinear Science and
  Numerical Simulation 126 (2023) 107458.
\newblock \href {https://doi.org/10.1016/j.cnsns.2023.107458}
  {\path{doi:10.1016/j.cnsns.2023.107458}}.

\bibitem{Brent:2002}
R.~Brent, Algorithms for Minimization without Derivatives, Dover Publications,
  Mineola, New York, 2002.

\bibitem{Broucke:1988}
R.~Broucke, The celestial mechanics of gravity assist, in: Proceedings of the
  AIAA/AAS Astrodynamics Conference, Minneapolis, Minnesota, August 15-17,
  1988, {Paper 88-4220-CP}.
\newblock \href {https://doi.org/10.2514/6.1988-4220}
  {\path{doi:10.2514/6.1988-4220}}.

\bibitem{Stough:2021}
R.~Stough, K.~Robinson, J.~Holt, D.~Smith, W.~Hitt, B.~Perry, Nasa's space
  launch system: Capabilities for ultra-high c3 missions, Bulletin of the
  American Astronomical Society 53~(4) (2021).
\newblock \href {https://doi.org/10.3847/25c2cfeb.170da7f3}
  {\path{doi:10.3847/25c2cfeb.170da7f3}}.

\end{thebibliography}

\end{document}